\documentclass[10pt,a4paper]{article}
\usepackage[utf8]{inputenc}
\usepackage[english]{babel}

\usepackage{amsmath}
\usepackage{amsfonts}
\usepackage{amssymb}

\RequirePackage[colorlinks,citecolor=blue,urlcolor=blue,linkcolor=blue]{hyperref}

\usepackage{authblk}

\usepackage{feynmp}
\usepackage[top=2cm,bottom=2cm]{geometry}

\newcommand{\pd}{\partial}

\begin{document}

\title{One-loop effective scalar-tensor gravity}

\author[1,2]{\href{https://orcid.org/0000-0001-7099-0861}{Boris Latosh}\footnote{\href{mailto:latosh@theor.jinr.ru}{latosh@theor.jinr.ru}}}

\affil[1]{Bogoliubov Laboratory of Theoretical Physics, JINR, Dubna 141980, Russia}
\affil[2]{Dubna State University, Universitetskaya str. 19, Dubna 141982, Russia}

\date{\today}

\maketitle

\begin{abstract}
  Non-minimal interactions are proven to be generated at the one-loop level in simple scalar-tensor gravity models. The John interaction from the Fab Four class is generated. The interaction affects the speed of gravitational waves in the contemporary Universe. Its role in low-energy phenomenology is discussed. Brans-Dicke-like interaction is generated in a non-minimal model. An opportunity to generate a dynamic low-energy Newton constant is addressed.
\end{abstract}

\section{Introduction}

Effective field theory framework provides a tool to study quantum effects in gravity models \cite{Georgi:1994qn,Donoghue:1994dn,Burgess:2003jk,Buchbinder:1992rb,Calmet:2013hfa}. Within the effective theory generated by general relativity some verifiable predictions were obtained. For instance, corrections for the Newton potential were studied \cite{BjerrumBohr:2002kt,Donoghue:1994dn,Akhundov:1996jd} together with PPN parameters \cite{Goldberger:2004jt,Levi:2018nxp}. Various implementations of effective theory for gravity models is widely covered in literature \cite{Calmet:2018qwg,Calmet:2019odl,Alexeyev:2017scq,Barvinsky:1985an,Bjerrum-Bohr:2014zsa,Kuntz:2019zef,Odintsov:1989sx,Odintsov:1989gz,Odintsov:1990qq}.

The effective theory for general relativity is constructed as follows \cite{Georgi:1994qn,Donoghue:1994dn,Burgess:2003jk,Buchbinder:1992rb,Calmet:2013hfa}. First of all, the theory is confined to an energy region below the Planck scale, as it marks the limit of applicability of general relativity. Secondly, a normalization scale $\mu$ is chosen below the Planck mass. At this scale a microscopic action $\mathcal{A}$ is defined. Finally, the theory is extended below the normalization scale via loop corrections and its description is given by an effective action $\Gamma$.

An effective theory constructed by this algorithm cannot be considered fundamental. The theory is confined to an energy region below the normalization scale $\mu$ which, in turn, is smaller then the Planck scale. There are no reasons to expect that the theory will be applicable outside this domain. A similar logic holds for the microscopic action $\mathcal{A}$. It can only be viewed as an approximation of the fundamental theory at $\mu$. The fundamental theory itself lies beyond the scope of the effective field theory framework. In such a way the framework allows one to study quantum effects without a detailed knowledge about the fundamental theory.

The microscopic action $\mathcal{A}$ defined at the normalization scale can be non-renormalizable within the standard quantum field theory. Firstly, as the theory is not fundamental there are no reasons to impose the renormalizability condition. Secondly, loop corrections can generate operators missing from the microscopic action $\mathcal{A}$. The corresponding infinite contributions can be safely normalized at the scale $\mu$ \cite{Donoghue:1994dn,Donoghue:2017pgk}. The fundamental theory, no matter if it will be renormalizable in the standard sense or not, should contain a suitable regularization mechanism. Therefore all divergences appearing in the effective theory will be regularized. The finite contributions generated by loop corrections can be recovered from empirical data. Finally, coefficients of non-analytic terms generated at the loop level can be evaluated within the effective theory itself \cite{Barvinsky:1985an}.

In the context of effective theory general relativity serves as a natural and the simplest gravity model to be defined at the normalization scale. However, it is not the only theory that can be used. Modified gravity models \cite{Berti:2015itd,Clifton:2011jh,Nojiri:2017ncd,Nojiri:2010wj} can generate effective models as well. Therefore, it is reasonable to expect that modifications of general relativity will modify the effective theory as well.

We address effective theories generated by simple scalar-tensor models which provide, perhaps, the simplest extension of general relativity. It may appear that a single additional scalar degree of freedom cannot influence an effective theory in a meaningful way, but this is not so. This claim is supported by previous studies \cite{Arbuzov:2017nhg,Latosh:2018xai} where it was shown that some non-minimal scalar-tensor models generate non-trivial modifications of the effective theory. Namely, at the one-loop level a new set of higher derivative operators is generated together with new interactions.

We show that the simplest scalar-tensor gravity generates a non-minimal scalar coupling to gravity at the one-loop level. The new interaction belongs to Horndeski gravity \cite{Horndeski:1974wa,Kobayashi:2011nu,Deffayet:2013lga}
, so the corresponding effective model may describe a significantly different low-energy phenomenology. We argue that it can be put to empirical verification via the latest data on gravitational wave speed \cite{Ezquiaga:2017ekz,TheLIGOScientific:2017qsa}. We also address a non-minimal scalar-tensor gravity with a simple scalar field self-interaction. In full agreement with the previous results \cite{Arbuzov:2017nhg} we show that in such a model a Brans-Dicke-like interaction is generated at the one-loop level. Consequently, the model develops an effective Newton constant with a scalar field dependence.

This paper is organized as follows. In Section \ref{section_the_simplest_model} we study the simplest effective scalar-tensor gravity. The model generated by general relativity with one additional massless scalar field with no self-interaction and with the minimal coupling to gravity. We show that at the one-loop level a non-minimal interaction with gravity is generated. In Section \ref{section_non-minimal_model} we address a scalar-tensor model with cubic and quartic self-interactions. We show that this model generates a Brans-Dicke-like interaction. We discuss possible implementations of these results in Section \ref{predictions}. We bring our conclusions in Section \ref{conclusions}.

\section{The simplest model}\label{section_the_simplest_model}

The simplest scalar-tensor gravity is given by the following action:
\begin{align}\label{the_first_microscopic_action}
  \mathcal{A} = \int d^4 x \sqrt{-g} \left[ - \cfrac{2}{\kappa^2}\, R +\cfrac12 ~g^{\mu\nu} \pd_\mu\phi \, \pd_\nu \phi \right].
\end{align}
Here $R$ is the scalar curvature, $\phi$ is the new scalar field, and $\kappa$ is related with the Newton constant $\kappa^2=32\pi\, G$. The action describes the simplest model as it has no scalar self-interaction and admits the minimal interaction with gravity. Such a setup allows one to obtain universal predictions. If a certain operator is generated by \eqref{the_first_microscopic_action}, then it will be generated in the most part of scalar-tensor models. Perhaps, the only exception for this observation can be given by models with non-dynamical scalar fields, but they present an exceptional case that lies beyond the scope of this discussion.

Following the standard effective field theory framework we define the microscopic action \eqref{the_first_microscopic_action} at some normalization scale $\mu$ below the Planck mass. At the normalization scale the model generates the following tree-level rules:
\begin{align*}
  \begin{gathered}
    \begin{fmffile}{R01}
      \begin{fmfgraph*}(30,30)
        \fmfleft{L}
        \fmfright{R}
        \fmf{dashes,label=$k$}{L,R}
      \end{fmfgraph*}
    \end{fmffile}
  \end{gathered}
  =& \cfrac{i}{k^2} ~, & 
  \begin{gathered}
    \begin{fmffile}{R02}
      \begin{fmfgraph*}(30,30)
        \fmfleft{L}
        \fmfright{R}
        \fmf{dbl_wiggly,label=$k$}{L,R}
        \fmflabel{$\mu\nu$}{L}
        \fmflabel{$\alpha\beta$}{R}
      \end{fmfgraph*}
    \end{fmffile}
  \end{gathered}\hspace{.7cm}
  =&\cfrac{i}{2} \cfrac{C_{\mu\nu\alpha\beta}}{k^2} ~, \\ \\
  \begin{gathered}
    \begin{fmffile}{R03}
      \begin{fmfgraph*}(30,30)
        \fmfleft{L}
        \fmfright{R1,R2}
        \fmf{dbl_wiggly}{L,V}
        \fmf{dashes}{R1,V}
        \fmf{dashes}{R2,V}
        \fmflabel{$p$}{R1}
        \fmflabel{$q$}{R2}
        \fmflabel{$\mu\nu$}{L}
        \fmfdot{V}
      \end{fmfgraph*}
    \end{fmffile}
  \end{gathered}
  =& i \,\cfrac{\kappa}{2} \,C^{\mu\nu\alpha\beta} p_{\alpha} q_\beta,  &
  \begin{gathered}
    \begin{fmffile}{R04}
      \begin{fmfgraph*}(30,30)
        \fmfleft{L1,L2}
        \fmfright{R1,R2}
        \fmf{dbl_wiggly}{L1,V}
        \fmf{dbl_wiggly}{L2,V}
        \fmf{dashes}{R1,V}
        \fmf{dashes}{R2,V}
        \fmflabel{$p$}{R1}
        \fmflabel{$q$}{R2}
        \fmflabel{$\mu\nu$}{L1}
        \fmflabel{$\alpha\beta$}{L2}
        \fmfdot{V}
      \end{fmfgraph*}
    \end{fmffile}
  \end{gathered}
  =& -i~2 ~\kappa^2 ~C_{(2)}^{\rho\sigma\mu\nu\alpha\beta} p_\rho q_\sigma ,
\end{align*}
\begin{align*}
  \begin{gathered}
    \begin{fmffile}{R05}
      \begin{fmfgraph*}(30,30)
        \fmfleft{L1,L2,L3}
        \fmfright{R1,R2}
        \fmf{dbl_wiggly}{L1,V}
        \fmf{dbl_wiggly}{L2,V}
        \fmf{dbl_wiggly}{L3,V}
        \fmf{dashes}{R1,V}
        \fmf{dashes}{R2,V}
        \fmflabel{$p$}{R1}
        \fmflabel{$q$}{R2}
        \fmflabel{$\mu\nu$}{L1}
        \fmflabel{$\alpha\beta$}{L2}
        \fmflabel{$\rho\sigma$}{L3}
        \fmfdot{V}
      \end{fmfgraph*}
    \end{fmffile}
  \end{gathered}
  =&-i~3!~\kappa^3 ~C_{(3)}^{\lambda\tau\mu\nu\alpha\beta\rho\sigma} p_\lambda q_\tau, \\
\end{align*}
\begin{align}\label{the_first_set_of_tree-level_rules}
  \begin{gathered}
    \begin{fmffile}{R06}
      \begin{fmfgraph*}(30,30)
        \fmftop{T1,T2}
        \fmfbottom{B}
        \fmf{dbl_wiggly}{B,V}
        \fmf{dbl_wiggly}{T1,V}
        \fmf{dbl_wiggly}{T2,V}
        \fmflabel{$\alpha\beta$,$q$}{T1}
        \fmflabel{$\rho\sigma$,$l$}{T2}
        \fmflabel{$\mu\nu$,$p$}{B}
        \fmfdot{V}
        \fmffreeze
      \end{fmfgraph*}
    \end{fmffile}
  \end{gathered}
  =& i \kappa ~ \big( T^{\mu\nu\alpha\beta\rho\sigma\lambda\tau} ~q_\lambda l_\tau +T^{\alpha\beta\mu\nu\rho\sigma\lambda\tau} ~p_\lambda l_\tau + T^{\rho\sigma\mu\nu\alpha\beta\lambda\tau} p_\lambda q_\tau \big). \\ \nonumber
\end{align}
The first two diagrams corresponds to propagators of particles carrying a momentum $k$. The graviton propagator is given in the harmonic gauge $\pd_\mu h^{\mu\nu}-1/2~\pd^\nu h =0$. From here on all momenta in all three-point diagrams are directed inwards. Definitions of tensors and a comment on a derivation of the rules are discussed in \ref{tensor_definitions}.

It is useful to address loop corrections to two point functions. One-loop corrections for the graviton propagator were studied in detail in \cite{tHooft:1974toh}. They generate higher derivative operators $R^2$ and $R_{\mu\nu}^2$. One-loop corrections to the scalar propagator, on the contrary, vanishes in $d=4$:
\begin{align}
  \begin{split}
    \begin{gathered}
      \begin{fmffile}{D00}
        \begin{fmfgraph}(45,45)
          \fmfleft{L}
          \fmfright{R}
          \fmf{phantom,tension=2}{L,l}
          \fmf{dashes,tension=.05}{l,r}
          \fmf{phantom,tension=2}{r,R}
          \fmf{dbl_wiggly,left=1}{l,r}
          \fmfdot{l,r}
        \end{fmfgraph}
      \end{fmffile}
    \end{gathered}
    =& -i~\cfrac{\kappa^2 p^4}{32}~ (d-2)(d-4) (2\pi\mu)^{4-d} \int\cfrac{d^dk}{(2\pi)^d}\cfrac{1}{k^2(k-p)^2} ~, \\
    \begin{gathered}
      \begin{fmffile}{D08}
        \begin{fmfgraph}(45,45)
          \fmfleft{L}
          \fmfright{R}
          \fmf{dashes}{L,V,R}
          \fmf{dbl_wiggly,right=1}{V,V}
          \fmfdot{V}
        \end{fmfgraph}
      \end{fmffile}
    \end{gathered}
    =&\cfrac{\kappa^2}{8}\, (3d-2)(d-4)\,p^2 (2\pi\mu)^{4-d}\,\int\cfrac{d^dk}{(2\pi)^d}\,\cfrac{1}{k^2} ~.
  \end{split}
\end{align}
Therefore no higher derivative operators are generated in the scalar sector of the effective theory and  it is free from the corresponding Ostrogradsky instability \cite{Ostrogradsky:1850fid,Woodard:2006nt}.

There are five one-loop amplitudes describing corrections to the scalar-graviton interaction:
\begin{align}\label{the_diagrams}
  \mathcal{M}_{(1)}{}_{\mu\nu}=&\hspace{.7cm}
  \begin{gathered}
    \begin{fmffile}{D01}
      \begin{fmfgraph*}(35,35)
        \fmfleft{L}
        \fmfright{R1,R2}
        \fmf{dbl_wiggly}{L,V}
        \fmf{dashes}{R1,D}
        \fmf{dashes}{U,R2}
        \fmf{dashes,tension=.5}{D,V,U}
        \fmf{dbl_wiggly,left=.5,tension=.5}{U,D}
        \fmflabel{$\mu\nu$}{L}
        \fmfdot{V,U,D}
      \end{fmfgraph*}
    \end{fmffile}
  \end{gathered} &
  \mathcal{M}_{(2)}{}_{\mu\nu}=&\hspace{.7cm}
  \begin{gathered}
    \begin{fmffile}{D02}
      \begin{fmfgraph*}(35,35)
        \fmfleft{L}
        \fmfright{R1,R2}
	\fmf{dbl_wiggly}{L,V}
        \fmf{dbl_wiggly,tension=.5,right=.5}{U,V,D}
        \fmf{dashes}{R1,D}
        \fmf{dashes}{U,R2}
        \fmf{dashes,tension=.5}{U,D}
        \fmflabel{$\mu\nu$}{L}
        \fmfdot{U,D,V}
      \end{fmfgraph*}
    \end{fmffile}
  \end{gathered} \nonumber \\
  \mathcal{M}_{(3)}{}_{\mu\nu}=&\hspace{.7cm}
  \begin{gathered}
    \begin{fmffile}{D03}
      \begin{fmfgraph*}(35,35)
        \fmfleft{L}
        \fmfright{R1,R2}
	\fmf{dbl_wiggly}{L,V}
        \fmf{dbl_wiggly,tension=.4,left=1}{V,VR,V}
        \fmf{dashes}{R1,VR,R2}
        \fmfdot{V,VR}
        \fmflabel{$\mu\nu$}{L}
      \end{fmfgraph*}
    \end{fmffile}
  \end{gathered} &
  \mathcal{M}_{(4)}{}_{\mu\nu}=&\hspace{.7cm}
  \begin{gathered}
    \begin{fmffile}{D04}
      \begin{fmfgraph*}(35,35)
        \fmfleft{L}
        \fmfright{R1,R2}
        \fmf{dbl_wiggly,tension=5}{L,V}
        \fmf{dashes}{R1,V,R2}
        \fmf{phantom}{R1,D}
        \fmf{phantom,tension=.3}{D,V}
        \fmffreeze
        \fmf{dbl_wiggly,left=.5}{V,D}
        \fmfdot{V,D}
        \fmflabel{$\mu\nu$}{L}
      \end{fmfgraph*}
    \end{fmffile}
  \end{gathered} \\
  \mathcal{M}_{(5)}{}_{\mu\nu}=&\hspace{.7cm}
  \begin{gathered}
    \begin{fmffile}{D05}
      \begin{fmfgraph*}(35,35)
        \fmfleft{L}
        \fmfright{R1,R2}
        \fmf{dbl_wiggly,tension=5}{L,V}
        \fmf{dashes,left=.5}{R1,V,R2}
        \fmf{phantom}{V,U}
        \fmf{phantom}{R1,U,R2}
        \fmffreeze
        \fmf{dbl_wiggly,left=1}{V,U,V}
        \fmfdot{V}
        \fmflabel{$\mu\nu$}{L}
      \end{fmfgraph*}
    \end{fmffile}
  \end{gathered} \nonumber
\end{align}
These amplitudes were discussed before in the context of potential interactions between gravity and matter \cite{Donoghue:1994dn,Akhundov:1996jd}. Namely, it was shown that only $\mathcal{M}_{(2)}$ and $\mathcal{M}_{(3)}$ generate power law corrections to the effective Newton potential \cite{Holstein:2004dn}.

We will show that amplitudes \eqref{the_diagrams} generate quasi-potential interactions (which depend on particles momenta) which are relevant for the low-energy phenomenology. Namely, they generate interaction $G^{\mu\nu}~ \pd_\mu\phi~ \pd_\nu\phi$ that belong to the Fab Four class of Horndeski gravity \cite{Charmousis:2011bf}. The interaction is relevant, as it affects the speed of gravitational waves in the contemporary Universe which is constrained via GW$170817$ data \cite{Ezquiaga:2017ekz,TheLIGOScientific:2017qsa}. We discuss this feature in more details in Section \ref{predictions}.

Let up proceed with the proof of existence of the new interaction. First of all, one should only study amplitudes \eqref{the_diagrams} with external scalars being fixed on-shell. As we are interested in the low-energy phenomenology, it is reasonable to discuss only real (on-shell) states of the scalar field. Graviton states, on the contrary, should not be fixed on-shell. This is due to the fact that the scalar field can interact with a regular matter via exchange of virtual gravitons. Because of this the corresponding amplitudes will contain traces of the new interaction with off-shell gravitons.

Secondly, not all divergent contributions are relevant within the proposed setup. Some amplitudes, for instance $\mathcal{M}_{(5)}$, contain quadratic divergences
\begin{align}
  \int\cfrac{d^4k}{(2\pi)^4}\cfrac{1}{k^2} ~.
\end{align}
The issue of quadratic divergences itself poses a fundamental problem related with the naturalness of a theory \cite{Jack:1990pz}. In the context of gravity the corresponding corrections can also be related with the value of the cosmological constant (i.e. with a non-vanishing vacuum expectation value of the gravitational field) \cite{Padilla:2015aaa,Zeldovich:1968ehl,Weinberg:1988cp}. A more detailed discussion of these issues lies far beyond the scope of the paper.

Corrections associated with quadratic divergences can hardly be relevant within effective theory. The corresponding divergent contributions should be normalized at $\mu$ together with any other divergent contributions. The quadratic divergence does not depend on any momenta, so it provides a universal contribution at all energy scales. To put it otherwise, it can only provide a finite shift to certain couplings. The values of all coupling, in turn, are normalized via empirical data at $\mu$. Therefore such finite shifts will be completely adsorbed by data defining the theory at the normalization scale. In such a way, despite the fact that quadratic divergences do present in the theory, within the proposed setup they are completely fixed by the normalization scale data.

Thirdly, as the external scalars are fixed on-shell, some contributions of \eqref{the_diagrams} vanish. This feature can be easily illustrated with $\mathcal{M}_{(4)}$. If momenta of the external scalars $p$ and $q$ are not fixed on-shell, then the amplitude reads (evaluated with FeynCalc $9.3.0$ \cite{Shtabovenko:2020gxv}, given up to terms with quadratic divergences):
\begin{align}
  \mathcal{M}_{(4)}{}_{\mu\nu} =&\cfrac{\kappa^3}{6}~p\cdot q ~ \left[p_\mu p_\nu -\cfrac14~p^2 \eta_{\mu\nu}\right] \int\cfrac{d^4k}{(2\pi)^4}\cfrac{1}{k^2 (k-p)^2} ~.
\end{align}
When the momentum $p$ of an external scalar is fixed on-shell $p^2=0$, the corresponding integral vanishes, so does the amplitude.

Such a setup allows one to study amplitudes $\mathcal{M}_{(1)}$, $\mathcal{M}_{(2)}$, and $\mathcal{M}_{(3)}$, as only they contain contributions relevant for the problem. The overall three-particle amplitude reads:
\begin{align}\label{the_first_amplitude}
  \begin{split}
    \mathcal{M}_{\mu\nu} =& \mathcal{M}_{(1)}{}_{\mu\nu} +\mathcal{M}_{(2)}{}_{\mu\nu}+\mathcal{M}_{(3)}{}_{\mu\nu} \\
    =& -\cfrac{\kappa^3}{2}~l^4 ~C_{\mu\nu\alpha\beta} p^\alpha q^\beta ~\int\cfrac{d^4k}{(2\pi)^4}\cfrac{1}{k^2 (k+p)^2(k-q)^2} ~ \\
    &+\cfrac{\kappa^3}{12} ~l^2~ \Bigg[ 19~C_{\mu\nu\alpha\beta} p^\alpha q^\beta +3 ~\eta_{\mu\nu} ~p\cdot q  +9~l^2~\left(\eta_{\mu\nu}-\cfrac{l_\mu l_\nu}{l^2}\right) \Bigg]~\int\cfrac{d^4k}{(2\pi)^4}\cfrac{1}{k^2 (k-l)^2} ~.
  \end{split}
\end{align}
Here $p$ and $q$ are momenta of the external on-shell scalars, $l=-p-q$ is an off-shell momentum of the graviton. 

The first term in \eqref{the_first_amplitude} is free from ultraviolet divergences. It only contains infrared divergences which can be regularized via soft graviton radiation, in full analogy with the standard quantum electrodynamics (which was noted back in \cite{Donoghue:1994dn}).

For the sake of completeness we would like to make a brief comment on the soft graviton radiation. The discussed integral is typical for quantum field models:
\begin{align}
  \int\cfrac{d^4 k}{(2\pi)^4}\,\cfrac{1}{k^2 \, (k+p)^2 \, (k-q)^2} \,.
\end{align}
For finite momenta $p$ and $q$ its behavior can be easily analyzed. In the ultraviolet sector ($p, q \ll k$) the influence of $p$ and $q$ can be neglected and the integral is simplified:
\begin{align}
  \int\limits_{k\gg p, q}\cfrac{d^4 k}{(2\pi)^4}\,\cfrac{1}{k\, (k+p)^2\, (k+q)^2} \sim \int\cfrac{d^4k}{(2\pi)^4} \cfrac{1}{\left(k^2\right)^3} \sim \Lambda^{-2}.
\end{align}
Here  $\Lambda$ is a cut-off scale. The right hand side of this expression vanishes in $\Lambda\to\infty$ limit, which shows that the integral is regular in the ultraviolet sector. The leading contribution at the infrared sector ($k \ll p, q$), on the contrary, is singular:
\begin{align}
  \int\limits_{k\ll p, q} \cfrac{d^4 k}{(2\pi)^4}\,\cfrac{1}{k^2\,(k+p)^2\,(k+q)^2}\sim\int\limits_{k\ll p,q}\,\cfrac{d^4k}{(2\pi)^4}\,\cfrac{1}{k^2 ~k\cdot p ~ k\cdot q}\,.
\end{align}
The right hand side of this expression is proportional to $\ln k$ which is singular in $\kappa\to 0$ limit.

This infrared singularity is regularized at the level of cross-sections via a soft photon radiation as follows. Any conceivable physical detector always has a finite precision on measurement. Therefore a given physical detector always has an infrared sensitivity threshold $\lambda_\text{detector}$ and it cannot detect gravitons carrying energy below $\lambda_\text{detector}$. Thus, in any physical experiment the following amplitude is measured:
\begin{align}
  \mathcal{M}_\text{observed}=
  \begin{gathered}
    \begin{fmffile}{S01}
      \begin{fmfgraph}(35,35)
        \fmfleft{L}
        \fmfright{R1,R2}
        \fmf{dbl_wiggly}{L,V}
        \fmf{dashes}{R1,V,R2}
        \fmfblob{10}{V}
      \end{fmfgraph}
    \end{fmffile}
  \end{gathered}
  +
  \begin{gathered}
    \begin{fmffile}{S02}
      \begin{fmfgraph}(35,35)
        \fmfleft{L}
        \fmfright{R1,R2,R3}
        \fmf{dbl_wiggly}{L,V}
        \fmf{dashes}{R1,V,R3}
        \fmfdot{V}
        \fmffreeze
        \fmf{phantom}{V,W,R3}
        \fmfdot{W}
        \fmffreeze
        \fmf{dbl_wiggly}{W,R2}
      \end{fmfgraph}
    \end{fmffile}
  \end{gathered} ~.
\end{align}
The first term of this amplitude corresponds to the processes \eqref{the_diagrams} discussed above. The second term accounts for processes describing a radiation of an on-shell gravitons with energies below $\lambda_\text{detector}$. Such real gravitons cannot be detected by a given physical detector, so all measurements will contain a contribution coming from such processes.

The square of such an amplitude $\lvert \mathcal{M}_\text{observed} \rvert^2$ is free from infrared divergences due to the structure of the integral. The square of the second term of $\mathcal{M}_\text{observed}$ contains the following contribution
\begin{align}
  \int\limits_{\lvert k \rvert < \lambda_\text{detector}} \cfrac{d^4 k}{(2\pi)^4}\,\cfrac{1}{k^2 (k+p)^2 (k-q)^2} \,.
\end{align}
Here the integration is performed up to the detection threshold $\lambda_\text{detector}$, as one only accounts for gravitons that cannot be detected with a given apparatus. This contribution matches the infrared sector of the discussed integral and completely compensates it. This allows one to cast out such infrared divergences away from any physical cross-sections. This approach to infrared divergences is well covered in multiple sources \cite{Peskin:1995ev,Itzykson:1980rh}.


Getting back to the amplitudes, the first term in \eqref{the_first_amplitude} generates a finite contribution proportional to the following operator:
\begin{align}
  &\kappa^2 ~l^2 ~C_{\mu\nu\alpha\beta} p^\alpha q^\beta\sim ~\kappa^3~ \square h^{\mu\nu} ~ C_{\mu\nu\alpha\beta} \pd^\alpha \phi~ \pd^\beta \phi\sim ~\kappa^2~ G^{\mu\nu} ~ \pd_\mu \phi~\pd_\nu\phi ~.
\end{align}
This operator should be included in the one-loop effective action which reads:
\begin{align}\label{the_first_effective_action}
  \Gamma = \int d^4 x\sqrt{-g} \Bigg[ -\cfrac{2}{\kappa^2} ~R +&\cfrac12~g^{\mu\nu}\pd_\mu\phi\, \pd_\nu\phi+\kappa^2\,\beta\, G^{\mu\nu}\,\pd_\mu \phi\,\pd_\nu\phi \Bigg] ~
\end{align}
with $\beta$ being a dimensionless constant. Here for the sake of simplicity we omitted new terms appearing in the gravity sector of the effective theory, as it is extensively discussed in other papers \cite{Donoghue:1994dn,Burgess:2003jk,Calmet:2013hfa,Barvinsky:1985an}.

Finally, the second term in \eqref{the_first_amplitude} has a divergent part which should be normalized at $\mu$. The corresponding non-analytic part is proportional to the log-function $\ln\left(-l^2/\mu^2\right)$ and it is singular in the infrared region. There are a few reasons to believe that the model is safe in the infrared region. Even simple massless models experience certain problems in the infrared sector \cite{Coleman:1973jx}. It is safe to assume that the infrared behavior of the model will be dynamically regularized in a way similar to \cite{Coleman:1973jx}. The size of the Universe can be used as the simplest regularization scale. The energy scale associated with the cosmological constant can also be used as a regularization parameter. Moreover, the value of the cosmological constant is related with loop corrections \cite{Padilla:2015aaa,Zeldovich:1968ehl,Weinberg:1988cp}, so it may very well serve as a suitable dynamical regularization mechanism. This reasoning provides grounds to believe that the model is safe in the infrared region. Therefore, in full analogy with the standard electrodynamics and previous papers on effective gravity \cite{Donoghue:1994dn,BjerrumBohr:2002kt,Bjerrum-Bohr:2014zsa}, corresponding corrections can be accounted via an introduction of new form-factors to the corresponding expression for the scalar-graviton vertex.

\section{Non-minimal model}\label{section_non-minimal_model}

The simplest way to extend model \eqref{the_first_microscopic_action} is to introduce cubic and quartic scalar field self-interactions. Study of the corresponding effective theory is completely similar to the previous case. Firstly, one defines the following microscopic action at the normalization scale $\mu$ below the Planck mass:
\begin{align}\label{the_second_microscopic_action}
  \mathcal{A}=\int d^4 x \sqrt{-g} \Bigg[ -\cfrac{2}{\kappa^2} ~R+&\cfrac12~g^{\mu\nu}~\pd_\mu\phi~\pd_\nu\phi +\cfrac{\lambda}{3!}~\phi^3 + \cfrac{g}{4!}~\phi^4 \Bigg] ~.
\end{align}
Here $\lambda$ is the cubic scalar coupling with a dimension of mass, $g$ is the dimensionless quartic scalar coupling. Secondly, one extends the model down to the low-energy regime. Thirdly, one evaluates one-loop amplitudes describing interaction of a graviton with two scalars.


The model \eqref{the_second_microscopic_action} extends the models \eqref{the_first_microscopic_action} discussed before, so it inherits the set of Feynman rules \eqref{the_first_set_of_tree-level_rules}. It should only be extended with two new rule describing scalar self-interaction:
\begin{align}
  \begin{gathered}
    \begin{fmffile}{R07}
      \begin{fmfgraph}(35,35)
        \fmfleft{L}
        \fmfright{R1,R2}
        \fmf{dashes}{L,V}
        \fmf{dashes}{R1,V}
        \fmf{dashes}{R2,V}
        \fmfdot{V}
      \end{fmfgraph}
    \end{fmffile}
  \end{gathered} &= i \,\lambda , &
  \begin{gathered}
    \begin{fmffile}{R08}
      \begin{fmfgraph}(35,35)
        \fmfleft{L1,L2}
        \fmfright{R1,R2}
        \fmf{dashes}{R1,V,R2}
        \fmf{dashes}{L1,V,L2}
        \fmfdot{V}
      \end{fmfgraph}
    \end{fmffile}
  \end{gathered} & = i\, g \,.
\end{align}

Consequently, only two new diagrams should be studied:
\begin{align}
  \mathcal{M}_{(6)}{}_{\mu\nu} =& \hspace{.7cm}
  \begin{gathered}
    \begin{fmffile}{D06}
      \begin{fmfgraph*}(30,30)
        \fmfleft{L}
        \fmfright{R1,R2}
        \fmf{dbl_wiggly,tension=2}{L,V}
        \fmf{dashes}{R1,D,V,U,R2}
        \fmffreeze
        \fmf{dashes,left=.4}{U,D}
        \fmflabel{$\mu\nu$}{L}
        \fmfdot{U,D,V}
      \end{fmfgraph*}
    \end{fmffile}
  \end{gathered} ~,
  &\mathcal{M}_{(7)}{}_{\mu\nu} =&\hspace{.7cm}
  \begin{gathered}
    \begin{fmffile}{D07}
      \begin{fmfgraph*}(30,30)
        \fmfleft{L}
        \fmfright{R1,R2}
        \fmf{dbl_wiggly,tension=2}{L,V}
        \fmf{dashes}{R1,D,V,U,R2}
        \fmf{phantom,tension=.01}{U,D}
        \fmf{phantom}{D,R1}
        \fmf{phantom}{U,R2}
        \fmffreeze
        \fmf{dashes,left=.4}{U,D}
        \fmf{dashes,right=.4}{U,D}
        \fmflabel{$\mu\nu$}{L}
        \fmfdot{U,D,V}
      \end{fmfgraph*}
    \end{fmffile}
  \end{gathered} ~.
\end{align}

In full analogy with the previous case, amplitude $\mathcal{M}_{(7)}$ is irrelevant due to the quadratic divergence. Amplitude $\mathcal{M}_{(6)}$ is given by the following expression:
\begin{align}\label{the_second_amplitude}
  \begin{split}
    \mathcal{M}_{(6)}{}_{\mu\nu} =& -\cfrac{\kappa\lambda^2}{2} ~C_{\mu\nu\alpha\beta}~p^\alpha q^\beta ~\int\cfrac{d^4k}{(2\pi)^4}\cfrac{1}{k^2 (k-q)^2 (k+p)^2} \\
    & -\cfrac{\kappa\lambda^2}{2} \cfrac{1}{l^2}~\left[ -3 ~C_{\mu\nu\alpha\beta} p^\alpha q^\beta - l^2 \left(  \eta_{\mu\nu} -\cfrac{l_\mu l_\nu}{l^2}\right)\right] \int\cfrac{d^4k}{(2\pi)^4}\cfrac{1}{k^2 (k+l)^2} ~.
  \end{split}
\end{align}
Here $p$ and $q$ are momenta of on-shell scalars, $l=-p-q$ is an off-shell momentum of the graviton. The structure of the amplitude is analogous to the previous case. The first term in \eqref{the_second_amplitude} is free from ultraviolet divergences. Its infrared divergences can be regularized via soft scalar radiation. Therefore the term generates a finite contribution proportional to the following operator
\begin{align}
  \lambda^2 ~\eta_{\mu\nu} \sim \kappa \lambda^2~ h~ \phi^2 \sim \lambda^2 ~R ~\phi^2 .
\end{align}
This interaction also belongs to Horndeski gravity and was found before \cite{Arbuzov:2017nhg}. We also would like to highlight that this non-minimal interaction is generated by the interaction odd in scalar field which were found to be important for renormalizability of certain models \cite{Barra:2019kda,Buchbinder:2019bcc}. Finally, the second term in \eqref{the_second_amplitude} can be treated in full analogy with the previous case. It should be normalized at the normalization scale and its influence should be accounted via an introduction of form factors. Its non-analytic part appears to be singular in the infrared region, but it is safe to assume that it can be protected from the singularity in full analogy with the previous case.

Therefore the effective action for the non-minimal model reads:
\begin{align}\label{the_second_effective_action}
  \begin{split}
    \Gamma = \int d^4x \sqrt{-g} \Bigg[& -\left(\cfrac{2}{\kappa^2}+\alpha\, \phi^2 \right)R + \cfrac12~g^{\mu\nu}\pd_\mu\phi~\pd_\nu\phi \\
      &+ \kappa^2\, \beta\, G^{\mu\nu}\pd_\mu\phi\,\pd_\nu\phi + \cfrac{1}{3!}~\lambda~\phi^3 + \cfrac{1}{4!}~g~\phi^4 \Bigg] ~.
  \end{split}
\end{align}
We discuss implications of these results in the next section.

\section{Implications for realistic scenarios}\label{predictions}

There are two main results presented in previous sections. Firstly, in the minimal model a non-minimal interaction $G^{\mu\nu}\,\pd_\mu\phi\,\pd_\nu\phi$ is generated. This interaction is known as the John interaction \cite{Charmousis:2011bf}. Secondly, in the non-minimal model a cubic scalar self-interaction generates a new Brans-Dicke-like interaction. These results have a series of corollaries that may be crucial for the low-energy phenomenology.

First of all, the John interaction is highlighted by its ability to screen an arbitrary cosmological constant on a cosmological background \cite{Charmousis:2011bf}. The interaction was studied in the classical regime and it was proven that a simple model admitting this interaction can consistently describe the late-time universe expansion \cite{Starobinsky:2016kua}. Moreover the action studied in \cite{Starobinsky:2016kua} matches with the effective action \eqref{the_first_effective_action}. The model is not free from disadvantages, as its perturbations become unstable in a proximity of the cosmological singularity. This disadvantage is relived within effective theory, as the theory itself cannot be applied in a proximity of the singularity.

The John interaction is generated in the minimal scalar-tensor model, so it is safe to assume that it will be generated universally. In particular, if the Higgs scalar is a true fundamental particle, then the John interaction should enter an effective theory based on the Standard Model. Because of this it is safe to use the effective model \eqref{the_first_effective_action} as the simplest effective model suitable for a description of the contemporary gravitational phenomenology.

At the same time, there is an indication of a possible inconsistency. Due to non-linear nature of the John interaction it changes the speed of tensor perturbations propagating about a cosmological background. These perturbations should be associated with gravitational waves propagating in the contemporary Universe and their speed is constrained by recent data \cite{Copeland:2018yuh,Ezquiaga:2017ekz,TheLIGOScientific:2017qsa}. In paper \cite{Ezquiaga:2017ekz} it is argued that the mere influence of the John interaction on the gravitational wave speed is enough to exclude it from any realistic model. On the other hand, in paper \cite{Copeland:2018yuh} it is argued that there is a loophole that may allow to make several scalar-tensor models consistent with the empirical data.

It is possible to establish an explicit numerical constraint on the model for it to be consistent with the latest data obtained by the LIGO collaboration \cite{Monitor:2017mdv}. Let us use an expression for the speed of tensor perturbations $c_t$ propagating over a cosmological background given in \cite{Kase:2018aps}:
\begin{align}
  c_t^2 = \cfrac{2 G_4 - (\dot\phi)^2 G_{5,\phi} - (\dot\phi)^2\,\ddot\phi\,G_{5,X}}{2 G_4 - 2 (\dot\phi)^2\,G_{4,X} + (\dot\phi)^2\, G_{5,\phi} - H \, (\dot\phi)^3\, G_{5,X}}.
\end{align}
Here $G_i$ are functions specifying the theory. The effective action \eqref{the_first_effective_action} is given by the following values of these functions (we adopt conventions of \cite{Kase:2018aps} for this particular calculation):
\begin{align}
  G_2 = X, ~G_3 =0, ~G_4 = \cfrac{1}{16\pi G}, ~ G_5 = - 32\pi G \beta \phi .
\end{align}
Therefore the speed for tensor perturbations reads:
\begin{align}
  c_t^2 &= \cfrac{1+ \beta\, (16\,\pi\, G\, \dot\phi)^2}{1- \beta\, (16\,\pi\, G\, \dot\phi)^2} = 1 + 2\, \beta\, (16\,\pi\, G \,\dot\phi^2) + O\left(G^4\right).
\end{align}
The data of the LIGO collaboration \cite{Monitor:2017mdv} provides the following constraint on a relation between the speed of gravitational waves $\nu_\text{GW}$ and the speed of electromagnetic waves $\nu_\text{EM}$:
\begin{align}
  -3 \times 10^{-15} \leq \cfrac{\nu_\text{GW} - \nu_\text{EM}}{\nu_\text{EM}} \leq + 7 \times 10^{-16}.
\end{align}
This constraint implies the following bounds on the effective model (at the leading order of $G^2$):
\begin{align}\label{The_main_constraint}
  -3 \times 10^{-15} \leq \left( 16\,\pi\, G\, \dot\phi\right)^2 \, \beta \leq +7 \times 10^{-16}.
\end{align}
Therefore an effective model \eqref{the_first_effective_action} (alongside the other similar models) is consistent with the current empirical data as long as \eqref{The_main_constraint} is satisfied.

This result points to two important consequences. Firstly, it shows that some models with a suppressed but non-vanishing John interaction can be consistent with the current empirical data on the gravitational waves speed. Equation \eqref{The_main_constraint} provides an explicit constraint on the model parameters for it to be consistent with the data. Secondly, claims made in papers \cite{Copeland:2018yuh,Ezquiaga:2017ekz} should be revisited, as the effective action \eqref{the_first_effective_action} provides an explicit example of a model containing the John interaction and consistent with the empirical data.

%

Secondly, the existence of a Brans-Dicke-like interaction should be addressed. We would like to note that the new non-minimal interaction \eqref{the_second_effective_action} is typical for models with conformal symmetry \cite{penrose_1964,AIHPA_1968__9_2_109_0}. Nonetheless we will refer to it as to Brans-Dicke-like interaction for the sake of simplicity.

The effective action \eqref{the_second_effective_action} is generated by a model with a non-vanishing scalar field potential, although only the cubic interaction generates a new gravitational coupling. At the same time, the scalar field develops an effective potential in full analogy with \cite{Coleman:1973jx}. A more detailed discussion of the effective potential lies beyond the scope of this paper, as it presents a separate independent problem, and as it appears at the next order in $\kappa$ (at the order of $\kappa^4$). Nonetheless, It is safe to assume that the scalar field develops a non-vanishing vacuum expectation value $\langle\phi\rangle$. This value contributes to the observed value of the low-energy effective Newton constant:
\begin{align}
  -\cfrac{2}{\kappa^2} R \to& \left( -\cfrac{2}{\kappa^2} +\alpha~\phi^2\right) R = -\cfrac{1}{16 \pi G_\text{eff}(\phi,\langle\phi\rangle)} ~R ~.\label{The_effective_Newton_constant}
\end{align}
A similar mechanism is well-known in gravity models \cite{Fujii:1974bq,Englert:1976ep,Minkowski:1977aj,Chudnovsky:1976zj,Matsuki:1977da,Smolin:1979ca,Linde:1979kf,Zee:1980sj,Nieh:1982nb,Adler:1982ri}. In such a way the effective Newton constant develops a dynamic dependence of the scalar field in an analogy with the Brans-Dicke theory.

In full analogy with the John coupling, it is possible to establish some explicit numerical constraints. The equation \eqref{The_effective_Newton_constant} provides a new expression for the observed Newton constant. This observed Newton constant can vary either due to a cosmological variations of $\langle\phi\rangle$ or due to large local fluctuations of the scalar field $\phi$. Local fluctuations of a scalar field can safely be neglected in any realistic scenario, as they are extremely hard to register. For instance, it requires the energy and preceision of LHC to detect local fluctuations of the Higgs scalar field. Therefore it is naturally to assume that any local fluctuations of the observed Newton constant would require similar effort to be detected. On the contrary, time variations of the observed Newton constant associated with the cosmological variation of the vacuum expectation value $\langle\phi\rangle$ can very well be constrained with a plethora of empirical data \cite{Copi:2003xd,Bambi:2005fi,Wu:2009zb,Gaztanaga:2001fh,Benvenuto:2004bs,Kaspi:1994hp,Thorsett:1996fr,DeglInnocenti:1995hbi,Pitjeva:2013xxa,Williams:2004qba,Hofmann:2018myc} (see also \cite{Uzan:2010pm,Vijaykumar:2020nzc} and the references therein). For the best of our knowledge, the best constraint is given by lunar laser ranging \cite{Hofmann:2018myc,Williams:2004qba}:
\begin{align}
  \left\lvert \cfrac{\dot G_\text{eff}}{G_\text{eff}} \right\rvert < (7.1\pm 7.6) \times 10^{-14} \text{ yr}^{-1}.
\end{align}
This provides the corresponding constraint (which holds up to $O(G^2)$ terms) on the scalar field vacuum expectation value variation in $c=\hbar=1$ unit:
\begin{align}
  \left\lvert 32 \pi \,G \, \alpha\, \phi\, \dot\phi \right\rvert < (7.1\pm 7.6) \times 10^{-14} \text{ yr}^{-1}.
\end{align}

A more detailed investigation of this phenomenon should be preformed, as the effective scalar field potential together with the scalar field vacuum expectation value are effected by loop corrections. Results of the investigation will be presented elsewhere, as it constitutes an independent problem. It is reasonable to anticipate the presence of a mechanism dynamically generating a new mass scale, in full analogy with \cite{Coleman:1973jx}. Therefore it is also required to understand its role within effective scalar-tensor gravity.

This section should be concluded as follows. Firstly, it should be understood if the John interaction generated at the one-loop level can be consistent with the empirical gravitational waves data. If a presence of a strongly suppressed but non-vanishing John interaction cannot be brought to an agreement with the empirical data, then certain conclusions about the structure of a microscopic action should be drawn. Secondly, the scalar sector of an effective theory should be investigated. The new Brans-Dicke-like interaction can lead to a generation of a dynamical Newton constant. The constant part of the generated constant will be defined by the effective scalar field potential induced by gravitational loop corrections. Investigation of these issues allows one to better understand the effective field theory implementation for scalar-tensor gravity.

\section{Conclusions}\label{conclusions}

In this paper we addressed one-loop effective theory for scalar-tensor gravity. We study two simple models. First one is the simplest model \eqref{the_first_microscopic_action} describing general relativity with one massless scalar field without self-interaction and with the minimal coupling to gravity. Second model \eqref{the_second_microscopic_action} describes general relativity with a massless scalar field that admits cubic and quartic self-interaction but that is coupled to gravity in the minimal way.

At the one-loop level the minimal model generates the John interaction from Horndeski gravity
\begin{align}
  \kappa^2 ~G^{\mu\nu}~\pd_\mu\phi ~\pd_\nu\phi ~.
\end{align}
The interaction is suppressed by the Planck mass squared, but it may influence the low-energy phenomenology. Small tensor perturbations of the metric propagating over a cosmological background should be associated with gravitational waves in an expanding universe. As it is discussed in Section \ref{predictions}, these perturbations are coupled to the scalar field via the John interaction. Because of this their speed can deviate from the speed of light \cite{Copeland:2018yuh,Ezquiaga:2017ekz,TheLIGOScientific:2017qsa}. Due to the non-linear nature of the interaction it may have a non-negligible influence on the gravitational wave speed despite being suppressed. Consequently, a further investigation is required to establish if a suppressed but non-vanishing John interaction is consistent with the known gravitational wave phenomenology. This investigation posses a separate problem that will be addressed elsewhere.

The non-minimal model generates a Brans-Dicke-like interaction at the one-loop level. The model admits a non-vanishing scalar field potential, so it is safe to assume that the scalar field develops a non-vanishing vacuum expectation value. As it is discussed in Section \ref{predictions}, the scalar field effective potential is also modified by loop corrections in full analogy with \cite{Coleman:1973jx}. This reasoning shows that the non-minimal model generates an effective low-energy Newton constant with a dynamical dependence on the scalar field. Therefore the value of the Newton constant observed in the low-energy regime can develop a finite shift defined by the non-vanishing vacuum expectation value of the scalar field. The one-loop effective scalar field potential generated in scalar-tensor gravity and its influence on the low-energy Newton constant will be studied in details elsewhere.

This brings us to the following conclusions. Scalar-tensor gravity provides, perhaps, the simplest alternative for general relativity. The low-energy phenomenology described by effective models powered by scalar-tensor gravity appears to be a more sophisticated subject. The John interaction generated at the one-loop level in the simplest model affects the speed of gravitational waves in an expanding Universe. The Brans-Dicke-like interaction affects the low-energy Newton constant and turns it into a dynamic quantity. These phenomena have no counterparts in the effective general relativity and they should be investigated further.

The role of the John interaction may provide an insight about the structure of a microscopic action generating the effective theory. If a strongly suppressed but non-vanishing John interaction is inconsistent with the known data on gravitational wave speed, then simple actions \eqref{the_first_microscopic_action} and \eqref{the_second_microscopic_action} cannot be used in realistic cases. Despite the fact that this case seems to be unlikely \cite{Copeland:2018yuh}, a numerical constraint on the John coupling should be obtained. The new Brans-Dicke-like interaction can change the role of the low-energy Newton constant. It can receive a finite shift due to a non-vanishing scalar field vacuum expectation value and to obtain a dynamical dependence from the scalar field. The value of the non-vanishing expectation value, in turn, will be affected by the one-loop effective scalar field potential in full analogy with \cite{Coleman:1973jx}. Therefore the role of the scalar sector of an effective model should be clarified.

\section*{Acknowledgements}

The work was supported by the Foundation for the Advancement of Theoretical Physics and Mathematics ``BASIS’’. The author would like to thanks A.B. Arbuzov for useful discussions.

\appendix

\section{Derivation of Feynman rules}\label{tensor_definitions}

For the sake of completeness we give a few comments about a derivation of Feynman rules used in the paper and present the set of tensors used in the paper.

First of all, the following standard tensors are used:
\begin{align}
  I_{\mu\nu\alpha\beta}=&\cfrac12\left(\eta_{\mu\alpha}\eta_{\nu\beta}+\eta_{\mu\beta}\eta_{\nu\alpha}\right) ,\nonumber \\
  C_{\mu\nu\alpha\beta}=&\eta_{\mu\alpha}\eta_{\nu\beta}+\eta_{\mu\beta}\eta_{\nu\alpha}-\eta_{\mu\nu}\eta_{\alpha\beta} .
\end{align}

Tensor $I_{\mu\nu\alpha\beta}$ serves as a unit tensor for symmetric rank-2 tensors. For an arbitrary symmetric tensor $T_{\mu\nu}$ the following holds:
\begin{align}
  T_{\mu\nu} = I_{\mu\nu\alpha\beta} T^{\alpha\beta}.
\end{align}
We use the following generalizations the unit tensor:
\begin{align}
  T^{\mu\sigma} T_\sigma^\nu =& I^{\mu\nu\alpha_1\beta_1\alpha_2\beta_2} T_{\alpha_1\beta_1} T_{\alpha_2\beta_2} , \nonumber \\
  T^{\mu\sigma}T_{\sigma\rho} T^{\rho\nu} =& I^{\mu\nu\alpha_1\beta_1\alpha_2\beta_2\alpha_3\beta_3} T_{\alpha_1\beta_1} T_{\alpha_2\beta_2} T_{\alpha_3\beta_3} .
\end{align}
Here $T_{\mu\nu}$ is an arbitrary symmetric tensor and $I$-tensors are given by the following formulae:
\begin{align}
  I_{\mu\nu\alpha_1\beta_1\alpha_2\beta_2}=& \cfrac18 \left(\eta_{\mu\alpha_1}\eta_{\nu\alpha_2}\eta_{\beta_1\beta_2} + \cdots \right) ,\\
  I_{\mu\nu\alpha_1\beta_1\alpha_2\beta_2\alpha_3\beta_3}=& \cfrac{1}{48}\left(\eta_{\mu\alpha_1}\eta_{\nu\alpha_2}\eta_{\beta_1\alpha_3}\eta_{\beta_2\beta_3} +\cdots \right) .\nonumber
\end{align}

Secondly, if the spacetime metric $g_{\mu\nu}$ describes small perturbations $h_{\mu\nu}$ over a flat background $\eta_{\mu\nu}$:
\begin{align}
  g_{\mu\nu} = \eta_{\mu\nu} + \kappa h_{\mu\nu},
\end{align}
then the following expansions hold:
\begin{align}
  \sqrt{-g}=&1+\cfrac{\kappa}{2}~ \eta^{\mu\nu} h_{\mu\nu} -\cfrac{\kappa^2}{8}~h^{\mu\nu} h^{\alpha\beta} C_{\mu\nu\alpha\beta} \\
  &+ \cfrac{\kappa^3}{48} ~ h^{\mu\nu} h^{\alpha\beta} h^{\rho\sigma} C_{\mu\nu\alpha\beta\rho\sigma} + O(\kappa^4), \nonumber\\
  \sqrt{-g} g^{\mu\nu} =& \eta^{\mu\nu} -\cfrac{\kappa}{2}~ C^{\mu\nu\alpha\beta} h_{\alpha\beta} + \kappa^2 C_{(2)}^{\mu\nu\alpha_1\beta_1\alpha_2\beta_2} h_{\alpha_1\beta_1}h_{\alpha_2\beta_2} \nonumber \\
  & +\kappa^3 C_{(3)}^{\mu\nu\alpha_1\beta_1\alpha_2\beta_2\alpha_3\beta_3} h_{\alpha_1\beta_1}h_{\alpha_2\beta_2}h_{\alpha_3\beta_3} +O(\kappa^4).\nonumber
\end{align}
Here the following tensors were used:
\begin{align}
  C_{\mu\nu\alpha\beta\rho\sigma}=&\cfrac83\Bigg[I_{\mu\nu\alpha\tau}I^{\lambda\tau}{}_{\alpha\beta\rho\sigma} + I_{\alpha\beta\lambda\tau}I^{\lambda\tau}{}_{\mu\nu\rho\sigma} + I_{\rho\sigma\lambda\tau} I^{\lambda\tau}{}_{\mu\nu\alpha\beta} \Bigg] -2\big[\eta_{\mu\nu} I_{\alpha\beta\rho\sigma}+\eta_{\alpha\beta}I_{\mu\nu\rho\sigma} \nonumber \\
      &+\eta_{\rho\sigma} I_{\mu\nu\alpha\beta} \big] +\eta_{\mu\nu}\eta_{\alpha\beta}\eta_{\rho\sigma} , \nonumber \\
    C_{(2)}^{\mu\nu\alpha_1\beta_1\alpha_2\beta_2}=&I^{\mu\nu\alpha_1\beta_1\alpha_2\beta_2}+\cfrac14\big[I^{\mu\nu\alpha_1\beta_1}\eta^{\alpha_2\beta_2} +I^{\mu\nu\alpha_2\beta_2}\eta^{\alpha_1\beta_1}\big]-\cfrac18~\eta^{\mu\nu} C^{\alpha_1\beta_1\alpha_2\beta_2} ,\nonumber \\
    C_{(3)}^{\mu\nu\alpha_1\beta_1\alpha_2\beta_2\alpha_3\beta_3}=&-I^{\mu\nu\alpha_1\beta_1\alpha_2\beta_2\alpha_3\beta_3} +\cfrac16\big[I^{\mu\nu\alpha_1\beta_1\alpha_2\beta_2}\eta^{\alpha_3\beta_3}+I^{\mu\nu\alpha_2\beta_2\alpha_3\beta_3}\eta^{\alpha_1\beta_1}+I^{\mu\nu\alpha_3\beta_3\alpha_1\beta_1}\eta^{\alpha_2\beta_2}\big] \nonumber \\
    &+\cfrac{1}{24}\big[I^{\mu\nu\alpha_1\beta_1}I^{\alpha_2\beta_2\alpha_3\beta_3} + I^{\mu\nu\alpha_2\beta_2}I^{\alpha_3\beta_3\alpha_1\beta_1}+I^{\mu\nu\alpha_3\beta_3}I^{\alpha_1\beta_1\alpha_2\beta_2}\big] \nonumber \\
    &+\cfrac{1}{48}~\eta^{\mu\nu} C^{\alpha_1\beta_1\alpha_2\beta_2\alpha_3\beta_3} .
\end{align}

These formulae are used to derive the tree-level rules describing scalar-graviton interaction:
\begin{align}
  \begin{split}
  &\int d^4 x ~\sqrt{-g}~g^{\mu\nu} ~\cfrac12 ~\pd_\mu \phi ~\pd_\nu\phi \\
  &=\int d^4 x \Big[ -\cfrac12~\phi \square \phi -\cfrac{\kappa}{4}~h_{\mu\nu} C^{\mu\nu\alpha\beta}~\pd_\alpha\phi~\pd_\beta\phi +\cfrac{\kappa^2}{2} ~h_{\mu\nu}h_{\alpha\beta}~C_{(2)}^{\mu\nu\alpha\beta\rho\sigma} ~\pd_\rho\phi~\pd_\sigma\phi  \\
      &\hspace{1.5cm}+\cfrac{\kappa^3}{2} ~h_{\mu\nu}h_{\alpha\beta}h_{\rho\sigma} ~C_{(3)}^{\mu\nu\alpha\beta\rho\sigma\lambda\tau} ~\pd_\lambda\phi\pd_\tau\phi + O(\kappa^4)\Big] .
  \end{split}
\end{align}

The structure of a cubic graviton interaction is recovered as follows. One should separate the complete derivative term in the general relativity action \cite{Novikov2006modern}:
\begin{align}
  \sqrt{-g} R = \sqrt{-g} g^{\mu\nu}&\left[\Gamma^\sigma_{\mu\rho} \Gamma^\rho_{\nu\sigma} -\Gamma^\sigma_{\mu\nu} \Gamma^\rho_{\sigma\rho} \right]+ \text{full derivative} .
\end{align}
The cubic graviton interaction is given by terms cubic in perturbations:
\begin{align}
  &-\cfrac{2}{\kappa^2}~\sqrt{-g} R \to -\cfrac{\kappa}{2}\Big[ h^{\mu\nu}\,\pd_\mu h_{\rho\sigma}\,\pd_\nu h^{\rho\sigma}-2\,h^{\mu\nu}\,\pd_\rho h_{\mu\sigma}\,\pd^\sigma h_\nu{}^\rho\nonumber \\
    &+2\,h^{\mu\nu}\,\pd_\rho h_{\mu\sigma}\,\pd^\rho h_\nu{}^\sigma -4\,h^{\mu\nu}\,\pd_\sigma h_{\rho\mu}\,\pd_\nu h^{\rho\sigma} +2\,h^{\mu\nu}\,\pd_\mu h_{\nu\sigma}\,\pd^\sigma h\nonumber \\
    &- h^{\mu\nu}\, \pd_\mu h\, \pd_\nu h +2\, h^{\mu\nu}\,\pd_\mu h\, \pd^\sigma h_{\nu\sigma} + 2\, h^{\mu\nu}\,\pd_\sigma h_{\mu\nu}\,\pd_\rho h^{\rho\sigma} \nonumber \\
    &- 2\,h^{\mu\nu}\,\pd^\sigma h_{\mu\nu}\, \pd_\sigma h -\cfrac12\,h\,\pd_\sigma h_{\mu\nu}\, \pd^\sigma h^{\mu\nu}+h\,\pd_\mu h_{\nu\sigma}\,\pd^\nu h^{\mu\sigma}\nonumber \\
    & -h\,\pd^\rho h_{\rho\sigma}\,\pd^\sigma h +\cfrac12\,h\,\pd_\sigma h\, \pd^\sigma h \Big] \nonumber \\
  =&\cfrac{\kappa}{4}~h_{\rho_1\sigma_1} \pd_\alpha h_{\rho_2\sigma_2} \pd_\beta h_{\rho_3\sigma_3} ~T^{\rho_1\sigma_1\rho_2\sigma_2\rho_3\sigma_3\alpha\beta} ~.
\end{align}
The tensor $T$ defining the structure of the interaction is obtained from the following tensor
\begin{align}\label{three-graviton_interaction_tensor}
  &\eta^{\mu\lambda}\eta^{\nu\tau} I^{\alpha\beta\rho\sigma}-2\,\eta^{\mu\alpha}\eta^{\nu\rho}\eta^{\lambda\sigma}\eta^{\tau\beta}+2\,\eta^{\mu\alpha}\eta^{\nu\rho}\eta^{\beta\sigma}\eta^{\lambda\tau}\nonumber \\
  &-4\,\eta^{\mu\beta}\eta^{\nu\tau}\eta^{\alpha\rho}\eta^{\lambda\sigma}+2\eta^{\mu\lambda}\eta^{\nu\alpha}\eta^{\beta\tau}\eta^{\rho\sigma}-\eta^{\mu\lambda}\eta^{\nu\tau}\eta^{\alpha\beta}\eta^{\rho\sigma}\nonumber\\
  &+2\,\eta^{\mu\lambda}\eta^{\nu\rho}\eta^{\alpha\beta}\eta^{\tau\sigma}+2\,I^{\mu\nu\alpha\beta}\eta^{\lambda\sigma}\eta^{\tau\rho} -2\, I^{\mu\nu\alpha\beta}\eta^{\lambda\tau}\eta^{\rho\sigma}\nonumber \\
  &-\cfrac12\,\eta^{\mu\nu}\eta^{\lambda\tau}I^{\alpha\beta\rho\sigma} +\eta^{\mu\nu}\eta^{\beta\sigma}\eta^{\lambda\rho}\eta^{\tau\alpha}-\eta^{\mu\nu}\eta^{\lambda\alpha}\eta^{\beta\tau}\eta^{\rho\sigma}\nonumber \\
  &+\cfrac12\,\eta^{\mu\nu}\eta^{\alpha\beta}\eta^{\rho\sigma}\eta^{\lambda\tau}
\end{align}
via symmetrization with respect to $\rho_1\leftrightarrow \sigma_1$, $\rho_2\leftrightarrow\sigma_2$, $\rho_3\leftrightarrow\sigma_3$, and $(\alpha,\rho_2,\sigma_2)\leftrightarrow(\beta,\rho_3,\sigma_3)$. In other words, $T$ respects the following relations:
\begin{align}
  T^{\rho_1\sigma_1\rho_2\sigma_2\rho_3\sigma_3\alpha\beta}=T^{\sigma_1\rho_1\rho_2\sigma_2\rho_3\sigma_3\alpha\beta} =T^{\rho_1\sigma_1\sigma_2\rho_2\rho_3\sigma_3\alpha\beta}=T^{\rho_1\sigma_1\rho_2\sigma_2\sigma_3\rho_3\alpha\beta}\nonumber \\
  =T^{\rho_1\sigma_1\rho_3\sigma_3\rho_2\sigma_2\beta\alpha}.\nonumber
\end{align}

A brief comment on the necessity to rederive the structure of the three-graviton vertex is due. For the best of our knowledge the corresponding Feynman rule was first explicitly presented in \cite{Donoghue:1994dn}, eqn (53). However, the expression contains an obvious misprint. In the 6th line of (53) \cite{Donoghue:1994dn} it is written $q^2 (I^{\sigma\mu}{}_{\alpha\beta} I_{\gamma\delta}{}_\sigma^\nu + I_{\alpha\beta}{}_\sigma^\nu I^{\sigma\mu}{}_{\alpha\delta}) $, while it should read $q^2 (I^{\sigma\mu}{}_{\alpha\beta} I_{\gamma\delta}{}_\sigma^\nu + I_{\alpha\beta}{}_\sigma^\nu I^{\sigma\mu}{}_{\gamma\delta}) $. The same misprint is repeated in a later paper \cite{Donoghue:2017pgk}. For the best of our knowledge the misprint is corrected \cite{Akhundov:1996jd}.

An independent derivation of the three-graviton interaction serves two purposes. First one is to present a more simple way to obtain such an expression, as aforementioned papers dismiss a detailed discussion of the calculations. Second one is to cast the expression in a more compact form \eqref{three-graviton_interaction_tensor} which is more suitable for a treatment with computer algebra packages such as FeynCalc \cite{Shtabovenko:2020gxv}.

\bibliographystyle{unsrturl}
\bibliography{One_loop_ST_gravity.bib}

\begin{thebibliography}{10}

\bibitem{Georgi:1994qn}
H.~Georgi.
\newblock {Effective field theory}.
\newblock {\em Ann. Rev. Nucl. Part. Sci.}, 43:209--252, 1993.
\newblock \href {https://doi.org/10.1146/annurev.ns.43.120193.001233}
  {\path{doi:10.1146/annurev.ns.43.120193.001233}}.

\bibitem{Donoghue:1994dn}
John~F. Donoghue.
\newblock {General relativity as an effective field theory: The leading quantum
  corrections}.
\newblock {\em Phys. Rev.}, D50:3874--3888, 1994.
\newblock \href {http://arxiv.org/abs/gr-qc/9405057}
  {\path{arXiv:gr-qc/9405057}}, \href
  {https://doi.org/10.1103/PhysRevD.50.3874}
  {\path{doi:10.1103/PhysRevD.50.3874}}.

\bibitem{Burgess:2003jk}
C.~P. Burgess.
\newblock {Quantum gravity in everyday life: General relativity as an effective
  field theory}.
\newblock {\em Living Rev. Rel.}, 7:5--56, 2004.
\newblock \href {http://arxiv.org/abs/gr-qc/0311082}
  {\path{arXiv:gr-qc/0311082}}, \href {https://doi.org/10.12942/lrr-2004-5}
  {\path{doi:10.12942/lrr-2004-5}}.

\bibitem{Buchbinder:1992rb}
I.~L. Buchbinder, S.~D. Odintsov, and I.~L. Shapiro.
\newblock {\em {Effective action in quantum gravity}}.
\newblock 1992.

\bibitem{Calmet:2013hfa}
Xavier Calmet.
\newblock {Effective theory for quantum gravity}.
\newblock {\em Int. J. Mod. Phys.}, D22:1342014, 2013.
\newblock \href {http://arxiv.org/abs/1308.6155} {\path{arXiv:1308.6155}},
  \href {https://doi.org/10.1142/S0218271813420145}
  {\path{doi:10.1142/S0218271813420145}}.

\bibitem{BjerrumBohr:2002kt}
N.~E.~J Bjerrum-Bohr, John~F. Donoghue, and Barry~R. Holstein.
\newblock {Quantum gravitational corrections to the nonrelativistic scattering
  potential of two masses}.
\newblock {\em Phys. Rev.}, D67:084033, 2003.
\newblock [Erratum: Phys. Rev.D71,069903(2005)].
\newblock \href {http://arxiv.org/abs/hep-th/0211072}
  {\path{arXiv:hep-th/0211072}}, \href
  {https://doi.org/10.1103/PhysRevD.71.069903, 10.1103/PhysRevD.67.084033}
  {\path{doi:10.1103/PhysRevD.71.069903, 10.1103/PhysRevD.67.084033}}.

\bibitem{Akhundov:1996jd}
Arif~A. Akhundov, S.~Bellucci, and A.~Shiekh.
\newblock {Gravitational interaction to one loop in effective quantum gravity}.
\newblock {\em Phys. Lett.}, B395:16--23, 1997.
\newblock \href {http://arxiv.org/abs/gr-qc/9611018}
  {\path{arXiv:gr-qc/9611018}}, \href
  {https://doi.org/10.1016/S0370-2693(96)01694-2}
  {\path{doi:10.1016/S0370-2693(96)01694-2}}.

\bibitem{Goldberger:2004jt}
Walter~D. Goldberger and Ira~Z. Rothstein.
\newblock {An Effective field theory of gravity for extended objects}.
\newblock {\em Phys. Rev.}, D73:104029, 2006.
\newblock \href {http://arxiv.org/abs/hep-th/0409156}
  {\path{arXiv:hep-th/0409156}}, \href
  {https://doi.org/10.1103/PhysRevD.73.104029}
  {\path{doi:10.1103/PhysRevD.73.104029}}.

\bibitem{Levi:2018nxp}
Michele Levi.
\newblock {Effective Field Theories of Post-Newtonian Gravity: A comprehensive
  review}.
\newblock 2018.
\newblock \href {http://arxiv.org/abs/1807.01699} {\path{arXiv:1807.01699}}.

\bibitem{Calmet:2018qwg}
Xavier Calmet and Boris Latosh.
\newblock {Three Waves for Quantum Gravity}.
\newblock {\em Eur. Phys. J.}, C78(3):205, 2018.
\newblock \href {http://arxiv.org/abs/1801.04698} {\path{arXiv:1801.04698}},
  \href {https://doi.org/10.1140/epjc/s10052-018-5707-2}
  {\path{doi:10.1140/epjc/s10052-018-5707-2}}.

\bibitem{Calmet:2019odl}
Xavier Calmet and Boris Latosh.
\newblock {The Spectrum of Quantum Gravity}.
\newblock 2019.
\newblock \href {http://arxiv.org/abs/1907.10024} {\path{arXiv:1907.10024}}.

\bibitem{Alexeyev:2017scq}
S.~O. Alexeyev, X.~Calmet, and B.~N. Latosh.
\newblock {Gravity induced non-local effects in the standard model}.
\newblock {\em Phys. Lett.}, B776:111--114, 2018.
\newblock \href {http://arxiv.org/abs/1711.06085} {\path{arXiv:1711.06085}},
  \href {https://doi.org/10.1016/j.physletb.2017.11.028}
  {\path{doi:10.1016/j.physletb.2017.11.028}}.

\bibitem{Barvinsky:1985an}
A.~O. Barvinsky and G.~A. Vilkovisky.
\newblock {The Generalized Schwinger-Dewitt Technique in Gauge Theories and
  Quantum Gravity}.
\newblock {\em Phys. Rept.}, 119:1--74, 1985.
\newblock \href {https://doi.org/10.1016/0370-1573(85)90148-6}
  {\path{doi:10.1016/0370-1573(85)90148-6}}.

\bibitem{Bjerrum-Bohr:2014zsa}
N.~E.~J. Bjerrum-Bohr, John~F. Donoghue, Barry~R. Holstein, Ludovic Plante, and
  Pierre Vanhove.
\newblock {Bending of Light in Quantum Gravity}.
\newblock {\em Phys. Rev. Lett.}, 114(6):061301, 2015.
\newblock \href {http://arxiv.org/abs/1410.7590} {\path{arXiv:1410.7590}},
  \href {https://doi.org/10.1103/PhysRevLett.114.061301}
  {\path{doi:10.1103/PhysRevLett.114.061301}}.

\bibitem{Kuntz:2019zef}
Adrien Kuntz, Federico Piazza, and Filippo Vernizzi.
\newblock {Effective field theory for gravitational radiation in scalar-tensor
  gravity}.
\newblock {\em JCAP}, 1905(05):052, 2019.
\newblock \href {http://arxiv.org/abs/1902.04941} {\path{arXiv:1902.04941}},
  \href {https://doi.org/10.1088/1475-7516/2019/05/052}
  {\path{doi:10.1088/1475-7516/2019/05/052}}.

\bibitem{Odintsov:1989sx}
S.~D. Odintsov.
\newblock {The Vilkovisky Effective Action in Quantum Gravity with SU(5) Grand
  Unification Theory}.
\newblock {\em Europhys. Lett.}, 10:287--292, 1989.
\newblock \href {https://doi.org/10.1209/0295-5075/10/4/001}
  {\path{doi:10.1209/0295-5075/10/4/001}}.

\bibitem{Odintsov:1989gz}
S.~D. Odintsov.
\newblock {The Parametrization Invariant and Gauge Invariant Effective Actions
  in Quantum Field Theory}.
\newblock {\em Fortsch. Phys.}, 38:371--391, 1990.

\bibitem{Odintsov:1990qq}
S.~D. Odintsov.
\newblock {Vilkovisky effective action in quantum gravity with matter}.
\newblock {\em Theor. Math. Phys.}, 82:45--51, 1990.
\newblock [Teor. Mat. Fiz.82,66(1990)].
\newblock \href {https://doi.org/10.1007/BF01028251}
  {\path{doi:10.1007/BF01028251}}.

\bibitem{Donoghue:2017pgk}
John~F. Donoghue, Mikhail~M. Ivanov, and Andrey Shkerin.
\newblock {EPFL Lectures on General Relativity as a Quantum Field Theory}.
\newblock 2017.
\newblock \href {http://arxiv.org/abs/1702.00319} {\path{arXiv:1702.00319}}.

\bibitem{Berti:2015itd}
Emanuele Berti et~al.
\newblock {Testing General Relativity with Present and Future Astrophysical
  Observations}.
\newblock {\em Class. Quant. Grav.}, 32:243001, 2015.
\newblock \href {http://arxiv.org/abs/1501.07274} {\path{arXiv:1501.07274}},
  \href {https://doi.org/10.1088/0264-9381/32/24/243001}
  {\path{doi:10.1088/0264-9381/32/24/243001}}.

\bibitem{Clifton:2011jh}
Timothy Clifton, Pedro~G. Ferreira, Antonio Padilla, and Constantinos Skordis.
\newblock {Modified Gravity and Cosmology}.
\newblock {\em Phys. Rept.}, 513:1--189, 2012.
\newblock \href {http://arxiv.org/abs/1106.2476} {\path{arXiv:1106.2476}},
  \href {https://doi.org/10.1016/j.physrep.2012.01.001}
  {\path{doi:10.1016/j.physrep.2012.01.001}}.

\bibitem{Nojiri:2017ncd}
S.~Nojiri, S.~D. Odintsov, and V.~K. Oikonomou.
\newblock {Modified Gravity Theories on a Nutshell: Inflation, Bounce and
  Late-time Evolution}.
\newblock {\em Phys. Rept.}, 692:1--104, 2017.
\newblock \href {http://arxiv.org/abs/1705.11098} {\path{arXiv:1705.11098}},
  \href {https://doi.org/10.1016/j.physrep.2017.06.001}
  {\path{doi:10.1016/j.physrep.2017.06.001}}.

\bibitem{Nojiri:2010wj}
Shin'ichi Nojiri and Sergei~D. Odintsov.
\newblock {Unified cosmic history in modified gravity: from F(R) theory to
  Lorentz non-invariant models}.
\newblock {\em Phys. Rept.}, 505:59--144, 2011.
\newblock \href {http://arxiv.org/abs/1011.0544} {\path{arXiv:1011.0544}},
  \href {https://doi.org/10.1016/j.physrep.2011.04.001}
  {\path{doi:10.1016/j.physrep.2011.04.001}}.

\bibitem{Arbuzov:2017nhg}
A.~B. Arbuzov and B.~N. Latosh.
\newblock {Fab Four self-interaction in quantum regime}.
\newblock {\em Eur. Phys. J.}, C77(10):702, 2017.
\newblock \href {http://arxiv.org/abs/1703.06626} {\path{arXiv:1703.06626}},
  \href {https://doi.org/10.1140/epjc/s10052-017-5233-7}
  {\path{doi:10.1140/epjc/s10052-017-5233-7}}.

\bibitem{Latosh:2018xai}
B.~Latosh.
\newblock {Fab Four Effective Field Theory Treatment}.
\newblock {\em Eur. Phys. J.}, C78(12):991, 2018.
\newblock \href {http://arxiv.org/abs/1812.01881} {\path{arXiv:1812.01881}},
  \href {https://doi.org/10.1140/epjc/s10052-018-6470-0}
  {\path{doi:10.1140/epjc/s10052-018-6470-0}}.

\bibitem{Horndeski:1974wa}
Gregory~Walter Horndeski.
\newblock {Second-order scalar-tensor field equations in a four-dimensional
  space}.
\newblock {\em Int. J. Theor. Phys.}, 10:363--384, 1974.
\newblock \href {https://doi.org/10.1007/BF01807638}
  {\path{doi:10.1007/BF01807638}}.

\bibitem{Kobayashi:2011nu}
Tsutomu Kobayashi, Masahide Yamaguchi, and Jun'ichi Yokoyama.
\newblock {Generalized G-inflation: Inflation with the most general
  second-order field equations}.
\newblock {\em Prog. Theor. Phys.}, 126:511--529, 2011.
\newblock \href {http://arxiv.org/abs/1105.5723} {\path{arXiv:1105.5723}},
  \href {https://doi.org/10.1143/PTP.126.511} {\path{doi:10.1143/PTP.126.511}}.

\bibitem{Deffayet:2013lga}
Cédric Deffayet and Danièle~A. Steer.
\newblock {A formal introduction to Horndeski and Galileon theories and their
  generalizations}.
\newblock {\em Class. Quant. Grav.}, 30:214006, 2013.
\newblock \href {http://arxiv.org/abs/1307.2450} {\path{arXiv:1307.2450}},
  \href {https://doi.org/10.1088/0264-9381/30/21/214006}
  {\path{doi:10.1088/0264-9381/30/21/214006}}.

\bibitem{Ezquiaga:2017ekz}
Jose~María Ezquiaga and Miguel Zumalacárregui.
\newblock {Dark Energy After GW170817: Dead Ends and the Road Ahead}.
\newblock {\em Phys. Rev. Lett.}, 119(25):251304, 2017.
\newblock \href {http://arxiv.org/abs/1710.05901} {\path{arXiv:1710.05901}},
  \href {https://doi.org/10.1103/PhysRevLett.119.251304}
  {\path{doi:10.1103/PhysRevLett.119.251304}}.

\bibitem{TheLIGOScientific:2017qsa}
B.~P. Abbott et~al.
\newblock {GW170817: Observation of Gravitational Waves from a Binary Neutron
  Star Inspiral}.
\newblock {\em Phys. Rev. Lett.}, 119(16):161101, 2017.
\newblock \href {http://arxiv.org/abs/1710.05832} {\path{arXiv:1710.05832}},
  \href {https://doi.org/10.1103/PhysRevLett.119.161101}
  {\path{doi:10.1103/PhysRevLett.119.161101}}.

\bibitem{tHooft:1974toh}
Gerard 't~Hooft and M.~J.~G. Veltman.
\newblock {One loop divergencies in the theory of gravitation}.
\newblock {\em Ann. Inst. H. Poincare Phys. Theor.}, A20:69--94, 1974.

\bibitem{Ostrogradsky:1850fid}
M.~Ostrogradsky.
\newblock {Mémoires sur les équations différentielles, relatives au
  problème des isopérimètres}.
\newblock {\em Mem. Acad. St. Petersbourg}, 6(4):385--517, 1850.

\bibitem{Woodard:2006nt}
Richard~P. Woodard.
\newblock {Avoiding dark energy with 1/r modifications of gravity}.
\newblock {\em Lect. Notes Phys.}, 720:403--433, 2007.
\newblock \href {http://arxiv.org/abs/astro-ph/0601672}
  {\path{arXiv:astro-ph/0601672}}, \href
  {https://doi.org/10.1007/978-3-540-71013-4_14}
  {\path{doi:10.1007/978-3-540-71013-4_14}}.

\bibitem{Holstein:2004dn}
Barry~R. Holstein and John~F. Donoghue.
\newblock {Classical physics and quantum loops}.
\newblock {\em Phys. Rev. Lett.}, 93:201602, 2004.
\newblock \href {http://arxiv.org/abs/hep-th/0405239}
  {\path{arXiv:hep-th/0405239}}, \href
  {https://doi.org/10.1103/PhysRevLett.93.201602}
  {\path{doi:10.1103/PhysRevLett.93.201602}}.

\bibitem{Charmousis:2011bf}
Christos Charmousis, Edmund~J. Copeland, Antonio Padilla, and Paul~M. Saffin.
\newblock {General second order scalar-tensor theory, self tuning, and the Fab
  Four}.
\newblock {\em Phys. Rev. Lett.}, 108:051101, 2012.
\newblock \href {http://arxiv.org/abs/1106.2000} {\path{arXiv:1106.2000}},
  \href {https://doi.org/10.1103/PhysRevLett.108.051101}
  {\path{doi:10.1103/PhysRevLett.108.051101}}.

\bibitem{Jack:1990pz}
I.~Jack and D.R.T. Jones.
\newblock Quadratic divergences and dimensional regularization.
\newblock {\em Nucl.Phys.B}, 342:127--148, 1990.
\newblock \href {https://doi.org/10.1016/0550-3213(90)90574-W}
  {\path{doi:10.1016/0550-3213(90)90574-W}}.

\bibitem{Padilla:2015aaa}
Antonio Padilla.
\newblock {Lectures on the Cosmological Constant Problem}.
\newblock 2015.
\newblock \href {http://arxiv.org/abs/1502.05296} {\path{arXiv:1502.05296}}.

\bibitem{Zeldovich:1968ehl}
{\relax Ya}.~B. Zel'dovich, Andrzej Krasinski, and {\relax Ya}.~B. Zeldovich.
\newblock {The Cosmological constant and the theory of elementary particles}.
\newblock {\em Sov. Phys. Usp.}, 11:381--393, 1968.
\newblock [Usp. Fiz. Nauk95,209(1968)].
\newblock \href {https://doi.org/10.1007/s10714-008-0624-6,
  10.1070/PU1968v011n03ABEH003927} {\path{doi:10.1007/s10714-008-0624-6,
  10.1070/PU1968v011n03ABEH003927}}.

\bibitem{Weinberg:1988cp}
Steven Weinberg.
\newblock {The Cosmological Constant Problem}.
\newblock {\em Rev. Mod. Phys.}, 61:1--23, 1989.
\newblock [,569(1988)].
\newblock \href {https://doi.org/10.1103/RevModPhys.61.1}
  {\path{doi:10.1103/RevModPhys.61.1}}.

\bibitem{Shtabovenko:2020gxv}
Vladyslav Shtabovenko, Rolf Mertig, and Frederik Orellana.
\newblock {FeynCalc 9.3: New features and improvements}.
\newblock 2020.
\newblock \href {http://arxiv.org/abs/2001.04407} {\path{arXiv:2001.04407}}.

\bibitem{Peskin:1995ev}
Michael~E. Peskin and Daniel~V. Schroeder.
\newblock {\em {An Introduction to quantum field theory}}.
\newblock Addison-Wesley, Reading, USA, 1995.

\bibitem{Itzykson:1980rh}
C.~Itzykson and J.B. Zuber.
\newblock {\em {Quantum Field Theory}}.
\newblock International Series In Pure and Applied Physics. McGraw-Hill, New
  York, 1980.

\bibitem{Coleman:1973jx}
Sidney~R. Coleman and Erick~J. Weinberg.
\newblock {Radiative Corrections as the Origin of Spontaneous Symmetry
  Breaking}.
\newblock {\em Phys. Rev.}, D7:1888--1910, 1973.
\newblock \href {https://doi.org/10.1103/PhysRevD.7.1888}
  {\path{doi:10.1103/PhysRevD.7.1888}}.

\bibitem{Barra:2019kda}
Vítor Fernandes~Barra, Iosif~L. Buchbinder, Jarme~Gomes Joaquim,
  Andreza~Rairis Rodrigues, and Ilya~L. Shapiro.
\newblock {Renormalization of Yukawa model with sterile scalar in curved
  spacetime}.
\newblock {\em Eur.\ Phys.\ J.\ C}, 79(6):458, 2019.
\newblock \href {http://arxiv.org/abs/1903.11546} {\path{arXiv:1903.11546}},
  \href {https://doi.org/10.1140/epjc/s10052-019-6917-y}
  {\path{doi:10.1140/epjc/s10052-019-6917-y}}.

\bibitem{Buchbinder:2019bcc}
Iosif~L. Buchbinder, Andreza~Rairis Rodrigues, Eduardo~Antonio dos Reis, and
  Ilya~L. Shapiro.
\newblock {Quantum aspects of Yukawa model with scalar and axial scalar fields
  in curved spacetime}.
\newblock {\em Eur.\ Phys.\ J.\ C}, 79(12):1002, 2019.
\newblock \href {http://arxiv.org/abs/1910.01731} {\path{arXiv:1910.01731}},
  \href {https://doi.org/10.1140/epjc/s10052-019-7447-3}
  {\path{doi:10.1140/epjc/s10052-019-7447-3}}.

\bibitem{Starobinsky:2016kua}
Alexei~A. Starobinsky, Sergey~V. Sushkov, and Mikhail~S. Volkov.
\newblock {The screening Horndeski cosmologies}.
\newblock {\em JCAP}, 1606(06):007, 2016.
\newblock \href {http://arxiv.org/abs/1604.06085} {\path{arXiv:1604.06085}},
  \href {https://doi.org/10.1088/1475-7516/2016/06/007}
  {\path{doi:10.1088/1475-7516/2016/06/007}}.

\bibitem{Copeland:2018yuh}
Edmund~J. Copeland, Michael Kopp, Antonio Padilla, Paul~M. Saffin, and
  Constantinos Skordis.
\newblock {Dark energy after GW170817 revisited}.
\newblock {\em Phys. Rev. Lett.}, 122(6):061301, 2019.
\newblock \href {http://arxiv.org/abs/1810.08239} {\path{arXiv:1810.08239}},
  \href {https://doi.org/10.1103/PhysRevLett.122.061301}
  {\path{doi:10.1103/PhysRevLett.122.061301}}.

\bibitem{Monitor:2017mdv}
B.P. Abbott et~al.
\newblock {Gravitational Waves and Gamma-rays from a Binary Neutron Star
  Merger: GW170817 and GRB 170817A}.
\newblock {\em Astrophys. J. Lett.}, 848(2):L13, 2017.
\newblock \href {http://arxiv.org/abs/1710.05834} {\path{arXiv:1710.05834}},
  \href {https://doi.org/10.3847/2041-8213/aa920c}
  {\path{doi:10.3847/2041-8213/aa920c}}.

\bibitem{Kase:2018aps}
Ryotaro Kase and Shinji Tsujikawa.
\newblock {Dark energy in Horndeski theories after GW170817: A review}.
\newblock {\em Int. J. Mod. Phys. D}, 28(05):1942005, 2019.
\newblock \href {http://arxiv.org/abs/1809.08735} {\path{arXiv:1809.08735}},
  \href {https://doi.org/10.1142/S0218271819420057}
  {\path{doi:10.1142/S0218271819420057}}.

\bibitem{penrose_1964}
R.~Penrose.
\newblock {\em Relativity, Groups and Topology}.
\newblock Gordon and Breach, 1964.

\bibitem{AIHPA_1968__9_2_109_0}
N.~A. Chernikov and E.~A. Tagirov.
\newblock Quantum theory of scalar field in de sitter space-time.
\newblock {\em Annales de l'I.H.P. Physique th\'eorique}, 9(2):109--141, 1968.
\newblock URL: \url{http://www.numdam.org/item/AIHPA_1968__9_2_109_0}.

\bibitem{Fujii:1974bq}
Y.~Fujii.
\newblock {Scalar-tensor theory of gravitation and spontaneous breakdown of
  scale invariance}.
\newblock {\em Phys. Rev.}, D9:874--876, 1974.
\newblock \href {https://doi.org/10.1103/PhysRevD.9.874}
  {\path{doi:10.1103/PhysRevD.9.874}}.

\bibitem{Englert:1976ep}
F.~Englert, C.~Truffin, and R.~Gastmans.
\newblock {Conformal Invariance in Quantum Gravity}.
\newblock {\em Nucl. Phys.}, B117:407--432, 1976.
\newblock \href {https://doi.org/10.1016/0550-3213(76)90406-5}
  {\path{doi:10.1016/0550-3213(76)90406-5}}.

\bibitem{Minkowski:1977aj}
Peter Minkowski.
\newblock {On the Spontaneous Origin of Newton's Constant}.
\newblock {\em Phys. Lett.}, 71B:419--421, 1977.
\newblock \href {https://doi.org/10.1016/0370-2693(77)90256-8}
  {\path{doi:10.1016/0370-2693(77)90256-8}}.

\bibitem{Chudnovsky:1976zj}
E.~M. Chudnovsky.
\newblock {The Spontaneous Conformal Symmetry Breaking and Higgs Model}.
\newblock {\em Theor. Math. Phys.}, 35:538, 1978.
\newblock [Teor. Mat. Fiz.35,398(1978)].
\newblock \href {https://doi.org/10.1007/BF01036453}
  {\path{doi:10.1007/BF01036453}}.

\bibitem{Matsuki:1977da}
Takayuki Matsuki.
\newblock {Effects of the Higgs Scalar on Gravity}.
\newblock {\em Prog. Theor. Phys.}, 59:235, 1978.
\newblock \href {https://doi.org/10.1143/PTP.59.235}
  {\path{doi:10.1143/PTP.59.235}}.

\bibitem{Smolin:1979ca}
Lee Smolin.
\newblock {Gravitational Radiative Corrections as the Origin of Spontaneous
  Symmetry Breaking!}
\newblock {\em Phys. Lett.}, 93B:95--100, 1980.
\newblock \href {https://doi.org/10.1016/0370-2693(80)90103-3}
  {\path{doi:10.1016/0370-2693(80)90103-3}}.

\bibitem{Linde:1979kf}
Andrei~D. Linde.
\newblock {Gauge theory and the variability of the gravitational constant in
  the early universe. (in russian)}.
\newblock {\em Pisma Zh. Eksp. Teor. Fiz.}, 30:479--482, 1979.
\newblock [Phys. Lett.93B,394(1980)].
\newblock \href {https://doi.org/10.1016/0370-2693(80)90350-0}
  {\path{doi:10.1016/0370-2693(80)90350-0}}.

\bibitem{Zee:1980sj}
A.~Zee.
\newblock {Spontaneously Generated Gravity}.
\newblock {\em Phys. Rev.}, D23:858, 1981.
\newblock \href {https://doi.org/10.1103/PhysRevD.23.858}
  {\path{doi:10.1103/PhysRevD.23.858}}.

\bibitem{Nieh:1982nb}
H.~T. Nieh.
\newblock {A Spontaneously Broken Conformal Gauge Theory Of Gravitation}.
\newblock {\em Phys. Lett.}, A88:388--390, 1982.
\newblock \href {https://doi.org/10.1016/0375-9601(82)90658-2}
  {\path{doi:10.1016/0375-9601(82)90658-2}}.

\bibitem{Adler:1982ri}
Stephen~L. Adler.
\newblock {Einstein Gravity as a Symmetry-Breaking Effect in Quantum Field
  Theory}.
\newblock {\em Rev. Mod. Phys.}, 54:729, 1982.
\newblock [Erratum: Rev. Mod. Phys.55,837(1983); ,539(1982)].
\newblock \href {https://doi.org/10.1103/RevModPhys.54.729}
  {\path{doi:10.1103/RevModPhys.54.729}}.

\bibitem{Copi:2003xd}
Craig~J. Copi, Adam~N. Davis, and Lawrence~M. Krauss.
\newblock {A New nucleosynthesis constraint on the variation of G}.
\newblock {\em Phys. Rev. Lett.}, 92:171301, 2004.
\newblock \href {http://arxiv.org/abs/astro-ph/0311334}
  {\path{arXiv:astro-ph/0311334}}, \href
  {https://doi.org/10.1103/PhysRevLett.92.171301}
  {\path{doi:10.1103/PhysRevLett.92.171301}}.

\bibitem{Bambi:2005fi}
Cosimo Bambi, Maurizio Giannotti, and F.L. Villante.
\newblock {The Response of primordial abundances to a general modification of
  G(N) and/or of the early Universe expansion rate}.
\newblock {\em Phys. Rev. D}, 71:123524, 2005.
\newblock \href {http://arxiv.org/abs/astro-ph/0503502}
  {\path{arXiv:astro-ph/0503502}}, \href
  {https://doi.org/10.1103/PhysRevD.71.123524}
  {\path{doi:10.1103/PhysRevD.71.123524}}.

\bibitem{Wu:2009zb}
Fengquan Wu and Xuelei Chen.
\newblock {Cosmic microwave background with Brans-Dicke gravity II: constraints
  with the WMAP and SDSS data}.
\newblock {\em Phys. Rev. D}, 82:083003, 2010.
\newblock \href {http://arxiv.org/abs/0903.0385} {\path{arXiv:0903.0385}},
  \href {https://doi.org/10.1103/PhysRevD.82.083003}
  {\path{doi:10.1103/PhysRevD.82.083003}}.

\bibitem{Gaztanaga:2001fh}
E.~Gaztanaga, E.~Garcia-Berro, J.~Isern, E.~Bravo, and I.~Dominguez.
\newblock {Bounds on the possible evolution of the gravitational constant from
  cosmological type Ia supernovae}.
\newblock {\em Phys. Rev. D}, 65:023506, 2002.
\newblock \href {http://arxiv.org/abs/astro-ph/0109299}
  {\path{arXiv:astro-ph/0109299}}, \href
  {https://doi.org/10.1103/PhysRevD.65.023506}
  {\path{doi:10.1103/PhysRevD.65.023506}}.

\bibitem{Benvenuto:2004bs}
Omar~G. Benvenuto, Enrique Garcia-Berro, and Jordi Isern.
\newblock {Asteroseismological bound on G/G from pulsating white dwarfs}.
\newblock {\em Phys. Rev. D}, 69:082002, 2004.
\newblock \href {https://doi.org/10.1103/PhysRevD.69.082002}
  {\path{doi:10.1103/PhysRevD.69.082002}}.

\bibitem{Kaspi:1994hp}
V.~M. Kaspi, J.~H. Taylor, and M.~F. Ryba.
\newblock {High - precision timing of millisecond pulsars. 3: Long - term
  monitoring of PSRs B1855+09 and B1937+21}.
\newblock {\em Astrophys. J.}, 428:713, 1994.
\newblock \href {https://doi.org/10.1086/174280} {\path{doi:10.1086/174280}}.

\bibitem{Thorsett:1996fr}
S.E. Thorsett.
\newblock {The Gravitational constant, the Chandrasekhar limit, and neutron
  star masses}.
\newblock {\em Phys. Rev. Lett.}, 77:1432--1435, 1996.
\newblock \href {http://arxiv.org/abs/astro-ph/9607003}
  {\path{arXiv:astro-ph/9607003}}, \href
  {https://doi.org/10.1103/PhysRevLett.77.1432}
  {\path{doi:10.1103/PhysRevLett.77.1432}}.

\bibitem{DeglInnocenti:1995hbi}
S.~Degl'Innocenti, G.~Fiorentini, G.G. Raffelt, B.~Ricci, and A.~Weiss.
\newblock {Time variation of Newton's constant and the age of globular
  clusters}.
\newblock {\em Astron. Astrophys.}, 312:345--352, 1996.
\newblock \href {http://arxiv.org/abs/astro-ph/9509090}
  {\path{arXiv:astro-ph/9509090}}.

\bibitem{Pitjeva:2013xxa}
E.V. Pitjeva and N.P. Pitjev.
\newblock {Relativistic effects and dark matter in the Solar system from
  observations of planets and spacecraft}.
\newblock {\em Mon. Not. Roy. Astron. Soc.}, 432:3431, 2013.
\newblock \href {http://arxiv.org/abs/1306.3043} {\path{arXiv:1306.3043}},
  \href {https://doi.org/10.1093/mnras/stt695}
  {\path{doi:10.1093/mnras/stt695}}.

\bibitem{Williams:2004qba}
James~G. Williams, Slava~G. Turyshev, and Dale~H. Boggs.
\newblock {Progress in lunar laser ranging tests of relativistic gravity}.
\newblock {\em Phys. Rev. Lett.}, 93:261101, 2004.
\newblock \href {http://arxiv.org/abs/gr-qc/0411113}
  {\path{arXiv:gr-qc/0411113}}, \href
  {https://doi.org/10.1103/PhysRevLett.93.261101}
  {\path{doi:10.1103/PhysRevLett.93.261101}}.

\bibitem{Hofmann:2018myc}
F.~Hofmann and J.~Müller.
\newblock {Relativistic tests with lunar laser ranging}.
\newblock {\em Class. Quant. Grav.}, 35(3):035015, 2018.
\newblock \href {https://doi.org/10.1088/1361-6382/aa8f7a}
  {\path{doi:10.1088/1361-6382/aa8f7a}}.

\bibitem{Uzan:2010pm}
Jean-Philippe Uzan.
\newblock {Varying Constants, Gravitation and Cosmology}.
\newblock {\em Living Rev. Rel.}, 14:2, 2011.
\newblock \href {http://arxiv.org/abs/1009.5514} {\path{arXiv:1009.5514}},
  \href {https://doi.org/10.12942/lrr-2011-2} {\path{doi:10.12942/lrr-2011-2}}.

\bibitem{Vijaykumar:2020nzc}
Aditya Vijaykumar, Shasvath~J. Kapadia, and Parameswaran Ajith.
\newblock {Constraints on the time variation of the gravitational constant
  using gravitational wave observations of binary neutron stars}.
\newblock 3 2020.
\newblock \href {http://arxiv.org/abs/2003.12832} {\path{arXiv:2003.12832}}.

\bibitem{Novikov2006modern}
S.P. Novikov and I.A. Taimanov.
\newblock {\em Modern Geometric Structures and Fields}.
\newblock Graduate studies in mathematics. American Mathematical Society, 2006.

\end{thebibliography}

\end{document}